\documentclass[apl,superscriptaddress,twocolumn]{revtex4-1}

\usepackage{times}
\usepackage{graphicx}
\begin{document}
\title{Room Temperature In-plane $\langle100\rangle$ Magnetic Easy Axis for Fe$_3$O$_4$/SrTiO$_3$(001):Nb Grown by Infrared PLD}

\author{Matteo Monti}
\author{Mikel Sanz}
\author{Mohamed Oujja}
\author{Esther Rebollar}
\author{Marta Castillejo}
\affiliation{Instituto de Qu\'{\i}mica F\'{\i}sica ``Rocasolano'', CSIC, Madrid E-28006, Spain}
\author{Francisco J. Pedrosa}
\author{Alberto Bollero}
\affiliation{IMDEA Nanociencia, Instituto Madrile\~{n}o de Estudios Avanzados en Nanociencia, Madrid E-28049, Spain}
\author{Julio Camarero}
\author{Jose Luis F. Cu\~{n}ado}
\affiliation{IMDEA Nanociencia, Instituto Madrile\~{n}o de Estudios Avanzados en Nanociencia, Madrid E-28049, Spain}
\affiliation{Universidad Aut\'{o}noma de Madrid}
\author{Norbert M. Nemes}
\affiliation{Dpto. de F\'{\i}sica Aplicada III, Universidad Complutense de Madrid, Madrid E-28040, Spain}
\author{Federico J. Mompean}
\author{Mar Garcia-Hern\'{a}ndez}
\affiliation{Instituto de Ciencia de Materiales de Madrid, CSIC, Madrid E-28049, Spain}
\author{Shu Nie}
\author{Kevin F. McCarty}
\affiliation{Sandia National Laboratories, Livermore, CA 94550, USA}
\author{Alpha T. N'Diaye}
\author{Gong Chen}
\author{Andreas K. Schmid}
\affiliation{Lawrence Berkeley National Laboratory, Berkeley, California 94720, USA}
\author{Jos\'{e} F. Marco}
\author{Juan de la Figuera}
\affiliation{Instituto de Qu\'{\i}mica F\'{\i}sica ``Rocasolano'', CSIC, Madrid E-28006, Spain}
\email{juan.delafiguera@iqfr.csic.es}

\begin{abstract}
We examine the magnetic easy-axis directions of stoichiometric magnetite films grown on SrTiO$_3$:Nb by infrared pulsed-laser deposition. Spin-polarized low-energy electron microscopy reveals that the individual magnetic domains are magnetized along the in-plane $\langle100\rangle$ film directions. Magneto-optical Kerr effect measurements show that the maxima of the remanence and coercivity are also along in-plane $\langle100\rangle$ film directions. This easy-axis orientation differs from bulk magnetite and films prepared by other techniques, establishing that the magnetic anisotropy can be tuned by film growth.

\end{abstract}

\maketitle

Magnetite (Fe$_3$O$_4$)\cite{CornellBook}, a ferrimagnet, is the oldest magnetic material known\cite{mills_lodestone_2004}. It is a highly correlated electron material that presents a prototypical metal-insulator transition close to 120~K (the Verwey transition\cite{walz_verwey_2002,garcia_verwey_2004}). At low temperature it becomes ferroelectric, and thus, multiferroic\cite{kato_observation_1982,alexe_ferroelectric_2009}. A bad metal at room temperature (RT), but predicted to be a half-metal with only the minority-spin band crossing the Fermi level\cite{katsnelson_half-metallic_2008}, it has been considered a promising material for spintronic applications as an spin-injector\cite{wada_efficient_2010} or as part of a spin-valve\cite{BibesAdvPhys2011}. For such purposes, it is often desired to obtain highly perfect magnetite films on different oxide substrates. In particular, SrTiO$_3$ is a very attractive material in the microelectronics industry and can be doped to provide either an insulating or metallic substrate. In consequence, there is interest in the magnetic and transport properties of magnetite films grown on SrTiO$_3$, both Nb-doped\cite{CarvelloJMMM2004,ZiesePRB2005,KundaliyaJAP2006,SatohSSC2008,WeiJPD2010} and undoped\cite{KalePRB2001,ZhengJVSTB2007,ChenJAP2008,ChengJCG2008,LeeSSC2008,LeungJCG2008,HamieThinSol2012}, by using techniques such as molecular beam epitaxy or pulsed-laser deposition (PLD).

The magnetization bulk easy-axis directions of Fe$_3$O$_4$ at RT are the cubic $\langle111\rangle$ ones. The first order anisotropy constant changes sign upon cooling to 130~K, temperature below which the easy axis are the $\langle100\rangle$ directions\cite{BickfordPR1950,AbeJPSJ1976,JacksonIRM2011}, down to Verwey transition at $\sim$120~K where the structure changes from cubic to monoclinic. Thus, in the (001) surface of bulk samples the magnetization is expected to lie along the projection of the bulk $\langle111\rangle$ on the (001) surface, i.e., the in-plane $\langle110\rangle$ directions\cite{WilliamsJGeophysRes1998}, an expectation confirmed by spin-polarized low-energy electron microscopy observations (SPLEEM)\cite{delaFigueraUltra2013}. Most magnetic studies of thin films on SrTiO$_3$ are performed by techniques such as magneto-optical Kerr effect (MOKE), and SQUID or vibrating-sample magnetometry (VSM), all of which average over the full thickness of the magnetite film\cite{KalePRB2001,ChengJCG2008,ChenJAP2008,WeiJPD2010,HamieThinSol2012}. In most cases, $\langle110\rangle$ in-plane directions are reported for the easy-axis\cite{KalePRB2001,BrandlmaierPRB2008,FoninJAP2011}, although some works indicate in-plane isotropic films\cite{ChengJCG2008}. There are several reports of real-space imaging of the surface domains by magnetic-force microscopy (MFM)\cite{ChenJAP2008,WeiJPD2010,HamieThinSol2012}, showing domains of about 60--100~nm in size, similar to the observed grain size, but they do not identify the local domain magnetization direction. On other (100) substrates, $\langle110\rangle$ easy-axis directions are also usually reported\cite{BloemenJMMM1998}. Although attempts have been made to modify the easy axis orientation by the use of piezoelectric substrates\cite{BrandlmaierPRB2008} or through growth on stepped substrates\cite{ShvetsPRB2008_uniaxial}, to our knowledge no four-fold $\langle100\rangle$ magnetization axis have been obtained in magnetite films.

\begin{figure}[hbt]
\centerline{\includegraphics[width=0.5\textwidth]{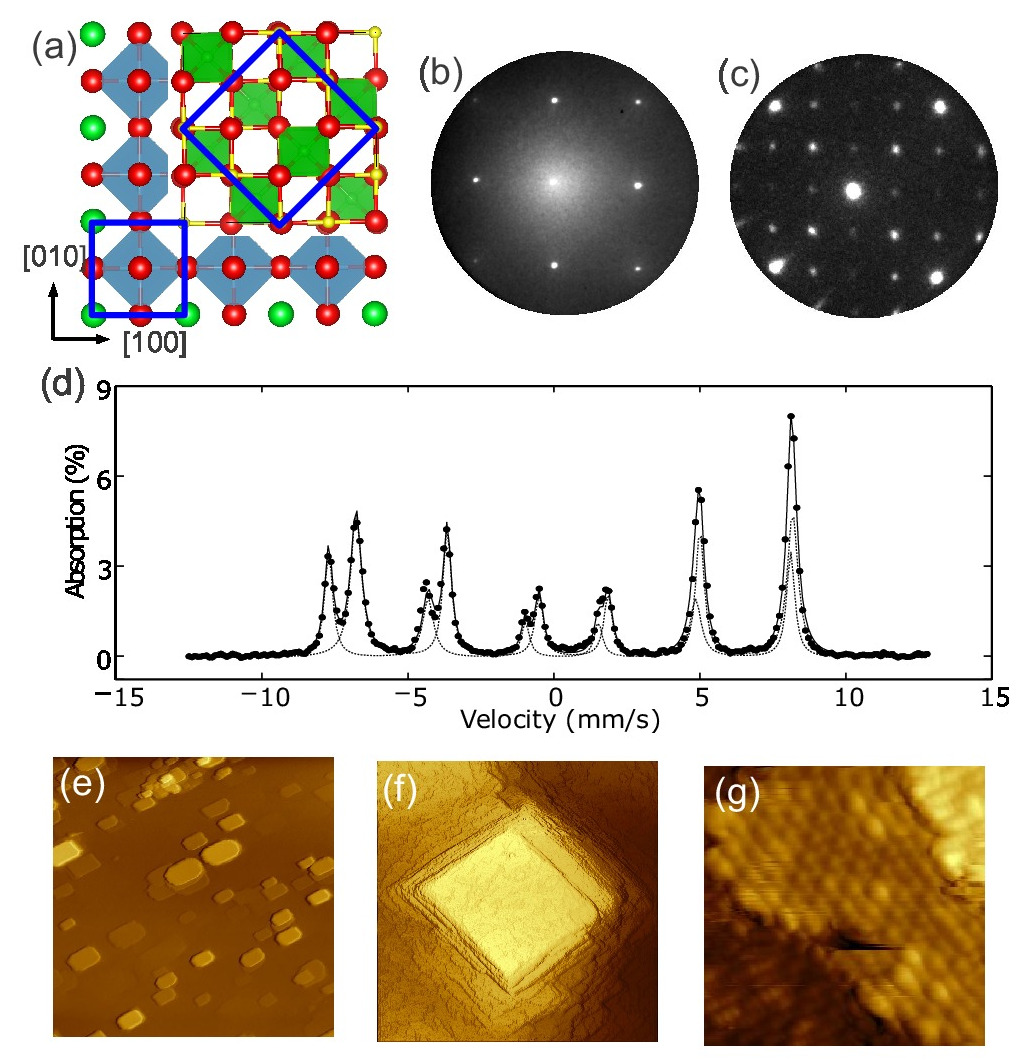}}
\caption{(a) Epitaxial relationship of magnetite on SrTiO$_3$. Oxygen atoms are shown as red spheres, with Sr atoms represented by green ones (Ti atoms are below in the middle of the blue-grey octahedra). The magnetite unit cell is shown in the upper-right side, with octahedral irons shown in yellow, and tetrahedral irons shown as green filled tetrahedra (schematics prepared by VESTA\cite{VESTA}). The surface unit cells of both materials are drawn by blue squares. (b) LEED pattern of SrTiO$_3$:Nb. (c) LEED pattern of the magnetite film grown on SrTiO$_3$:Nb by PLD. The LEED patterns have been acquired in LEEM at electron energies of 30 and 24~eV respectively (we note that in LEEM the area sampled in reciprocal space does not change with electron energy so both images are at the same scale). (d) ICEMS M\"{o}ssbauer spectra. (e) AFM image, 2~$\mu$m wide. The thermal color height scale corresponds to 40~nm. (f) STM image acquired with I$_t$=0.7~nA and V$_{bias}$=1~V. The image is 200~nm wide, and 10~nm high. (g) STM image acquired with I$_t$=0.5~nA and V$_{bias}$=1.1~V. The image is 41~nm wide, and shows the rows of atoms with different orientation in consecutive terraces. }
\label{microscopy_and_diffraction}
\end{figure} 

In this work we report on the growth by {\em infrared} pulsed-laser deposition (PLD)\cite{MikelASS2013} of highly perfect magnetite films on SrTiO$_3$:Nb and their characterization by a variety of techniques. The films present robust in-plane four-fold easy axes at RT but, in contrast with precedent results, they are oriented along the $\langle100\rangle$ directions as detected locally by SPLEEM and averaged by MOKE.

As for many complex oxides\cite{MoussyJPd2013}, one of the preferred growth methods for magnetite on SrTiO$_3$ has been PLD. In contrast to previous reported work using ultraviolet light, we have grown magnetite films by infrared PLD at 1064~nm using a hematite target\cite{MikelASS2013}. The Q-switched Nd:YAG laser had a full width at half-maximum of 15 ns with a 10 Hz repetition rate at a typical fluence of 4~J/cm$^2$. SrTiO$_3$(100) substrates doped with 0.1\% Nb from Crystek were heated to 780~K during deposition. Data reported in this work is from films 160~nm thick, although similar results have been obtained in 50~nm thick ones.

The epitaxial relationship between the perovskite substrate and the spinel film is expected to be  cube-on-cube, $[100]_{\mathrm Fe_3O_4} \parallel [100]_{\mathrm SrTiO_3}$, and $[010]_{\mathrm Fe_3O_4} \parallel [010]_{\mathrm SrTiO_3}$ depicted in Figure~\ref{microscopy_and_diffraction}(a).  The low-energy electron diffraction (LEED) pattern for SrTiO$_3$:Nb (after annealing the substrate in 10$^{-6}$~Torr of O$_2$ at 800~K for several hours) is shown in Figure~\ref{microscopy_and_diffraction}(b). The surface showed parallel steps $\sim100$~nm apart in LEEM and AFM (not shown). The magnetite LEED pattern measured after several cycles of cleaning by Ar$^+$ sputtering and annealing in 10$^{-6}$ Torr O$_2$, typical for preparing a clean magnetite surface in ultra-high vacuum\cite{ParkinsonSS2011}, is shown in Figure~\ref{microscopy_and_diffraction}(c). The strongest spots correspond to the first order and second order unit vectors of the surface primitive cell, which is rotated by 45$^\circ$ relative to the fcc cubic-cell unit vectors. The additional diffracted beams correspond to the $c(2\times2)$ reconstruction of magnetite\cite{PentchevaSS2008,ParkinsonSS2011}. Thus, the LEED pattern confirms the cube-of-cube epitaxial relationship, and it further shows that the film is monocrystalline.

The films have been characterized by integral conversion electron M\"{o}ssbauer spectroscopy (ICEMS), x-ray diffraction (XRD), x-ray photoemission spectroscopy (XPS) and SQUID magnetometry. In an ICEMS RT spectrum of stoichiometric magnetite, two sextet components are detected corresponding to iron in the octahedral and tetrahedral positions, respectively\cite{deGraveHI2000}. In such spectrum the component corresponding to octahedral iron presents parameters intermediate of those of Fe$^{2+}$ and Fe$^{3+}$. The magnetite film spectrum shown in Figure~\ref{microscopy_and_diffraction}(d) has been fitted with two components that have the expected values for magnetite, in isomer shift (0.23~mm/s and 0.69~mm/s), quadrupole shift (-0.04~mm/s and 0.01~mm/s) and hyperfine magnetic fields (49.0 T and 46.4 T). The ratio of the two components is 1.8--1.9 depending on the particular sample indicating that the film is of stoichiometric composition. The out-of-plane lattice spacing from XRD is 0.840~nm, indicating that the magnetite film is mostly relaxed. This is expected given the 7.5\% mismatch between magnetite and SrTiO$_3$. Our films thickness (50-160~nm) is well above the limit for pseudomorhic growth\cite{KalePRB2001} as detected by TEM\cite{ZhengJVSTB2007,HamieThinSol2012}. No contamination was detected by XPS, which showed only Fe and O, the former corresponding to a typical magnetite spectra\cite{deGrootPRB1999} with a mixture of Fe$^{2+}$ and Fe$^{3+}$. The Verwey temperature was measured to be 114~K with a SQUID magnetometer.

Typical microscopy images of the films are presented in Figure~\ref{microscopy_and_diffraction}(e--g). AFM images of the surface show square features (``mesas''), with heights of up to 30~nm and lateral sizes in the 100-200~nm range, emerging from a flat film, as shown in Figure~\ref{microscopy_and_diffraction}(e). In STM, both the areas between the mesas and their tops are confirmed to be quite flat, with small terraces tens of nanometers wide separated by atomic steps [0.21~nm high, see Figure~\ref{microscopy_and_diffraction}(f) where the contrast has been enhanced so individual steps can be located]. While the orientation of the atomic steps, both on top of the mesas and on the areas between them, is not well defined, the mesas themselves are remarkably well aligned with the in-plane $\langle110\rangle$ directions. On the individual atomic terraces, atomic rows 0.6~nm apart run along the $[110]$ direction in one terrace, and along the $[1\bar{1}0]$ direction of the next atomic terrace [Figure~\ref{microscopy_and_diffraction}(g)]. These rows correspond to the octahedral rows of iron of the magnetite unit cell, see Figure~\ref{microscopy_and_diffraction}(a)\cite{ParkinsonSS2011}. Along the rows there is also an additional 0.6~nm periodicity, out-of-phase between consecutive rows. These periodicities corresponds to the $c(2\times2)$ reconstruction observed by LEED shown in Figure~\ref{microscopy_and_diffraction}(c). This reconstruction, typical of magnetite cleaned by cycles of Ar$^+$ sputtering and annealing in vacuum, has been interpreted as a Jahn-Teller distortion of the topmost octahedral iron atom positions along the rows of the surface\cite{PentchevaSS2008}.

Cleaning the sample for ultra-high vacuum experiments (i.e., for the STM and LEEM observations), which involve mild sputtering and annealing, changes slightly the as-grown surface morphology. While it is obvious that individual atomic step positions are changed, we remark that the AFM measurements were done on the ``as-grown'' films. The agreement between the STM, LEEM and AFM results indicates that no large morphological changes have occurred during UHV cleaning.  

\begin{figure}
\centerline{\includegraphics[width=0.5\textwidth]{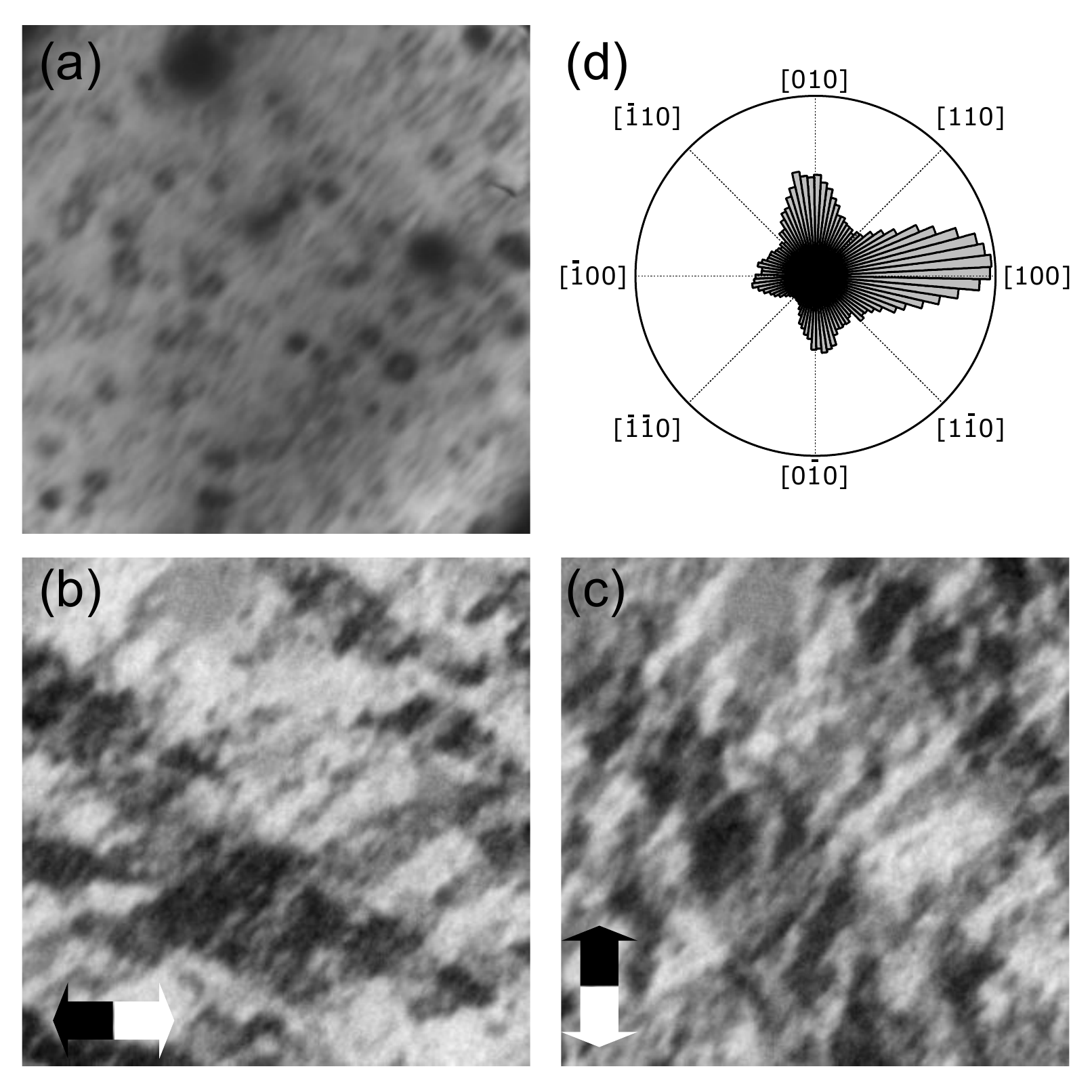}}
\caption{(a) RT LEEM image of a magnetite film on SrTiO$_3$:Nb.  (b,c) SPLEEM images acquired at the same location as (a) with the electron spin-direction along the x-axis ([100] direction) and y axis ($[0\bar{1}0]$) direction respectively showing the local surface magnetization component along the given direction. The LEEM start voltage for all the images is 6.8~V. The images are 9.1~$\mu$m wide. (d) Polar histogram of the in-plane magnetization as deduced from the images shown in (b,c).}
\label{SPLEEM}
\end{figure}

We have imaged the magnetic domains by means of SPLEEM\cite{RougemailleEPJAP2010}. After cleaning the samples with Ar$^+$ sputtering and annealing, the film was heated above the Curie temperature and slowly cooled back to RT. A low-energy electron micrograph of the film is shown in Figure~\ref{SPLEEM}(a). Faint squares are observed, which correspond to the mesas detected in the AFM and STM images in Figure~\ref{microscopy_and_diffraction}. The squares are oriented along in-plane $\langle110\rangle$ providing an internal direction reference. The two magnetic contrast images are obtained by calculating the pixel-by-pixel asymmetry between LEEM images acquired illuminating the sample with beams of electrons with opposite spin polarization: bright (dark) areas indicate that the local surface magnetization has a component parallel (anti-parallel) to the spin-polarization direction of the electron beam. Grey areas indicate the absence of a magnetization component along the spin-polarization direction. The SPLEEM images thus indicate the local surface magnetization along a given direction. We note that SPLEEM is extremely surface sensitive, detecting the magnetization of the topmost atomic layers of the film. As the electron beam spin-polarization can be changed with respect to the sample, the magnetization vector can be determined in real space with nanometer resolution\cite{Ramchal2004PRB}. More details on the SPLEEM instrument\cite{grzelakowski_new_1994}, the spin-polarization control method\cite{duden_compact_1995} or the vector magnetometric application of SPLEEM can be found in the literature\cite{Ramchal2004PRB,FaridPRL2006,FaridNJP2008}. SPLEEM images acquired (not shown) with out-of-plane spin direction presented negligible contrast, indicating that the magnetization lies mostly within the film plane. In Figures~\ref{SPLEEM}(b) and (c) white, black and grey regions are easily resolved. As the domains are not very large, it is difficult to determine by visual inspection whether the magnetization lies along some preferred axis. A plot of the magnetization-vector histogram vs. angle is shown in Figure~\ref{SPLEEM}(d) obtained by combining the images pixel by pixel to calculate the in-plane magnetization vector. It indicates that the magnetization lies mostly along the four in-plane $\langle100\rangle$ directions, i.e. the histogram shows peaks at angles corresponding to the [100], [010], $[\bar{1}00]$ and $[0\bar{1}0]$ orientations. The domain walls also show a preferred orientation, but along [110] and $[1\bar{1}0]$ directions, i.e. along the sides of the 3D mesas on the film. The domains are up to one micrometer in size.

\begin{figure}
\centerline{\includegraphics[width=0.5\textwidth]{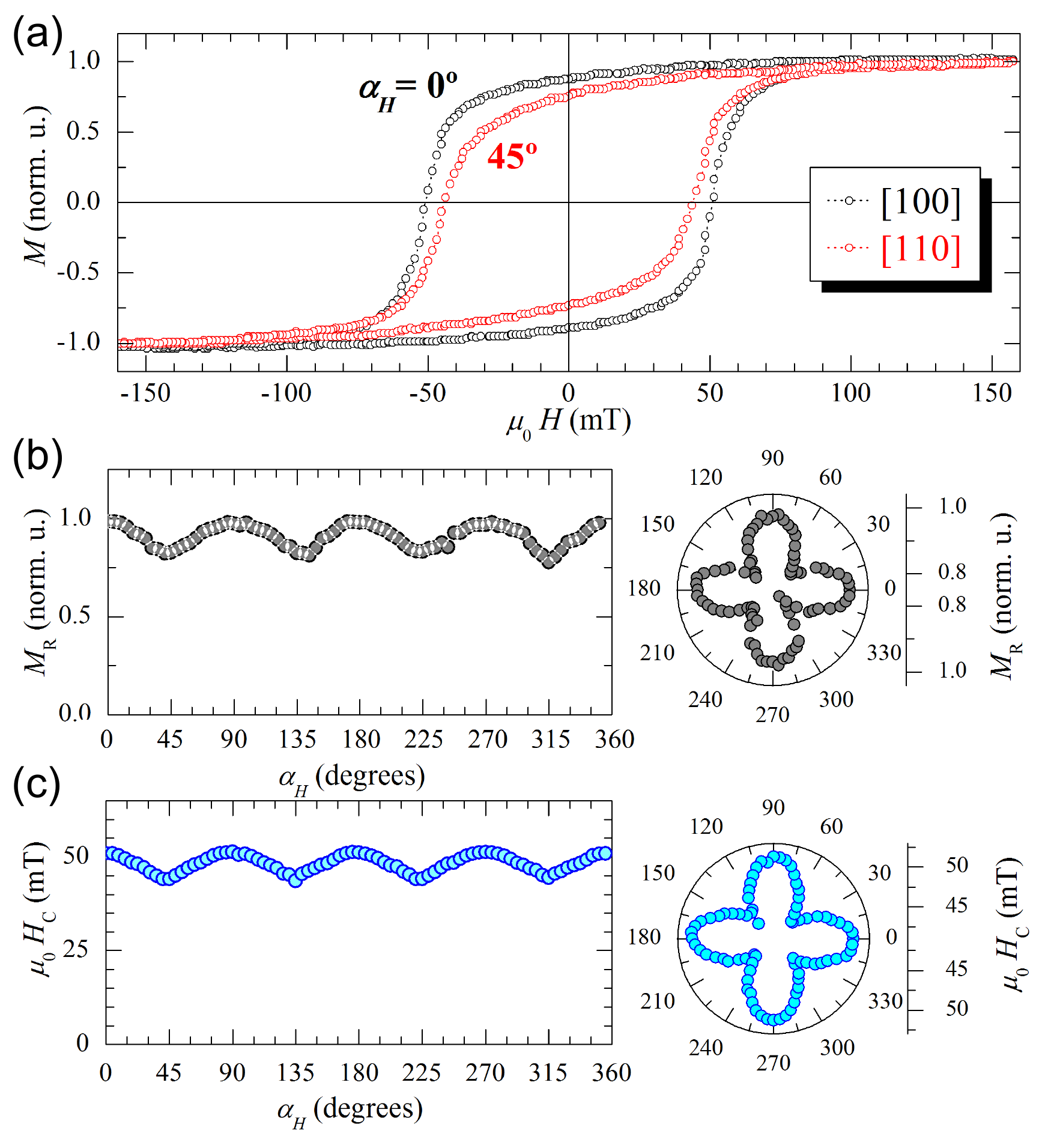}}
\caption{(a) RT hysteresis cycles acquired at $\alpha_H = 0^\circ$ ([100] direction) and at $\alpha_H = 45^\circ$ ([110] direction). (b) Angular evolution of the remanence, $M_R$, with corresponding polar-plot representation (right-hand side). (c) Angular evolution of the coercivity, $H_C$, with corresponding polar-plot representation (right).}
\label{MOKE}
\end{figure}

The unexpected magnetization easy-axis orientation of the film is confirmed by angular-dependent averaged magnetometry measurements. MOKE hysteresis loops have been systematically recorded by changing the in-plane orientation of the applied magnetic field in the angular range $\alpha_H$ = 0--360$^\circ$ (Figure~\ref{MOKE}). Two representative plots ($\alpha_H$ = 0$^\circ$ and 45$^\circ$) are shown in Figure~\ref{MOKE}(a). The film coercivity is 45--50 mT, in line with published values for magnetite on SrTiO$_3$\cite{ChenJAP2008,ChengJCG2008,WeiJPD2010}. The low saturation fields around 150~mT evidence the high structural quality of our films, although often much higher saturation fields are reported for magnetite films on 100 substrates, probably due to magnetic domain pinning at defects\cite{BerkowitzPRL1997}. Hysteresis loops displayed in Figure~\ref{MOKE}(a) show different remanence and coercivity values as a function of the in-plane orientation of the applied magnetic field.  In particular, larger remanence and coercivity values are found for $\alpha_H=0^\circ$, i.e., along the [100] direction. Moreover, The angular dependence of the remanence [Figure~\ref{MOKE}(b)] and the coercivity [Figure~\ref{MOKE}(c)] show the fourfold symmetry of the magnetic anisotropy with the highest values found at 0, 90, 180 and 270$^\circ$. The fourfold symmetry can be easily identified in the corresponding polar plots of the remanence and the coercivity [right-hand side of Figures~\ref{MOKE}(b) and (c)]. As the maxima of both remanence and coercivity correspond to to easy-axis directions, the angular MOKE indicates that for the full film the easy-axes are the in-plane $\langle100\rangle$ directions, in agreement with the microscopic SPLEEM observations of the surface magnetization.

In summary, we have grown pure stoichiometric magnetite films on SrTiO$_3$:Nb by infrared PLD. Unlike films reported to date, these films have a robust well defined easy-axis along the in-plane $\langle100\rangle$ directions. The individual magnetic domains at the surface of the films have been imaged in remanence by SPLEEM. The magnetic domains present magnetization vectors along the in-plane $\langle100\rangle$ directions, while the domain walls are aligned with the in-plane $\langle110\rangle$ directions. Hysteresis cycles have been measured by MOKE, obtaining the angular dependence of remanence and coercivity which both have maxima at the $\langle100\rangle$ directions and thus confirm the local determination of the easy-axis directions by an averaging technique. Our films prove that modifying the growth parameters in magnetite films allows tunning the easy-axis directions. 

\section*{Acknowledgments}
Authors acknowledge fruitful discussions with Prof. M. Ziese. This research was supported by Projects CTQ2010-15680, MAT2009-14578-C03-01 (MICINN), MAT2012-38045-C04-01 (MINECO), MAT2011-27470-C02-02 (MICINN), MAT2011-25598 (MINECO), the EU-FP7 NANOPYME Project (No. 310516) and by the Office of Basic Energy Sciences, Division of Materials and Engineering Sciences, U.~S. Department of Energy under Contract No. DE-AC04-94AL85000 (Sandia National Laboratories). Experiments performed at the National Center for Electron Microscopy, Lawrence Berkeley National Laboratory, were supported by the Office of Science, Office of Basic Energy Sciences, Scientific User Facilities Division, of the U.S. Department of Energy under Contract No. DE-AC02—05CH11231. E.R, M.O., M.S., A.T.N. and M.M. gratefully thank financial support  from the Ram\'{o}n y Cajal Programme (MINECO), a CSIC contract, a Geomateriales (CAM, S2009/Mat-1629) contract, a Feodor Lynen Postdoctoral Fellowship from the Alexander von Humboldt Foundation and a contract through the MICINN FPI Programme, respectively. We are grateful to Prof. T. Ezquerra (IEM, CSIC) for the use of the AFM system and M. Juanco (ICA, CSIC) for XRD measurements.

\bibliography{magnetite_on_sto}

\end{document}